\documentclass[aps, prl, twocolumn, showpacs, preprintnumbers, amssymb, floatfix]{revtex4}
\usepackage{amsmath}
\usepackage{amsfonts}
\usepackage{amsbsy}
\usepackage{xcolor}
\usepackage{soul}
\usepackage{graphicx}

\newcommand{\spvec}[1]{\ensuremath{\mathbf{#1}}}
\newcommand{\unitvec}[1]{\ensuremath{\mathbf{\hat{#1}}}}
\newcommand{\sptensor}[1]{\ensuremath{\boldsymbol{\mathbf{#1}}}}

\newcommand{\cmatrix}[1]{\ensuremath{\mathrm{#1}}}
\newcommand{\colvec}[1]{\ensuremath{\mathrm{#1}}}

\newcommand{\indMat}{\ensuremath{\mathfrak{L}}} 

\newcommand{\commentout}[1]{{}}
\newcommand{\half}{\hbox{$1\over2$}}
\newcommand{\eq}[1]{Eq.~\eqref{#1}}
\newcommand{\rv}{\spvec{r}}
\newcommand{\Ev}{\spvec{E}}
\newcommand{\Dv}{\spvec{D}}
\newcommand{\Bv}{\spvec{B}}
\newcommand{\Hv}{\spvec{H}}
\newcommand{\Pv}{\spvec{P}}

\newcommand{\pol}{\unitvec{e}}
\newcommand{\qv}{\spvec{q}}
\newcommand{\kv}{\spvec{k}}

\newcommand{\beq}{\begin{equation}}
\newcommand{\eeq}{\end{equation}}

\begin{document}
\title{Collective electromagnetic response of discrete metamaterial systems}
\author{S. D. Jenkins}
\author{J. Ruostekoski}
\affiliation{School of Mathematics and Centre for Photonic
  Metamaterials, University of Southampton,
  Southampton SO17 1BJ, United Kingdom}
\begin{abstract}
  We develop a general formalism to describe the interactions between
  a discrete set of plasmonic resonators mediated by the
  electromagnetic field.
  The resulting system of equations for closely-spaced meta-atoms
  represents a cooperative metamaterial response and we find that
  collective interactions between asymmetric split-ring resonators
  arranged in a 2D lattice  explains the recent experimental
  observations of system-size dependent narrowing of the transmission
  resonance linewidth.
  We further show that this cooperative narrowing depends sensitively
  on the lattice spacing.
\end{abstract}

\date{\today}
\pacs{42.25.Bs,41.20.Jb,31.10.-f}
\maketitle

Resonant multiple scattering plays an important role in mesoscopic
wave phenomena that can also be reached with electromagnetic (EM)
fields.
In the strong scattering regime, interference of different scattering
paths between discrete scatterers can result in, e.g.,  light
localization \cite{wiersma}--an effect analogous to the Anderson localization of
electrons in solids.
Metamaterials comprise artificially structured media of plasmonic
resonators interacting with EM fields.
Due to several promising phenomena, such as the possibility for
diffraction-free lenses resulting from negative refractive index \cite{negativeindex},
there has been a rapidly increasing interest in fabrication and
theoretical modeling of such systems.
The role of strong scattering and cooperative effects between
\emph{discrete} resonators in metamaterial arrays are largely unknown,
as such phenomena cannot be easily captured by commonly employed
theoretical techniques based on the homogenization approximation of
the scattering medium or full solutions of Maxwell's equations in
small, computationally accessible, systems.

In this letter we develop a theoretical formalism of collective
interactions between plasmonic resonators, or meta-atoms, mediated by
EM field.
In a computationally efficient model, which also provides physical
insight into the interaction processes, 
we
treat each meta-atom as a discrete scatterer, exhibiting a single mode
of current oscillation and possessing appropriate electric and
magnetic dipole moments.
Interactions with the EM field then determine the collective
interactions within the ensemble, resulting in collective resonance
frequencies, linewidths, and line shifts.
The resulting set of equations can represent a strong cooperative response in the
case of sufficiently closely-spaced resonators, necessitating the inclusion of
interference effects in multiple scattering between the meta-atoms.
As a specific example, we study asymmetric split-ring (ASR)
resonators, each consisting of two circular arcs of slightly unequal
lengths.
The currents in these ASRs may be excited symmetrically
(antisymmetrically), yielding a net oscillating electric (magnetic)
dipole.
In a single ASR we derive eigenstates analogous to superradiant and
subradiant states in a pair of atoms and in an ensemble of ASRs we
demonstrate how strong collective interactions in a discrete
metamaterial array are responsible for recent experimental
observations of a dramatic narrowing of the transmission resonance
linewidth (increasing quality factor) with the number of resonators~\cite{FedotovEtAlPRL2010}.
Our analysis indicates the necessity of accounting for the strong
collective response of metamaterial systems in understanding the
dynamics and design of novel meta-materials.
Strong interactions between resonators can find important applications
in metamaterial systems, providing, e.g., precise control and manipulation of
EM fields on a sub-wavelength
scale~\cite{SentenacPRL2008} and disorder-related
phenomena~\cite{papasimakis2009}.

We consider an ensemble of $N$ metamaterial unit elements,
metamolecules, each formed by $n$ discrete meta-atoms, such that the
position of the meta-atom $j$ is denoted by $\spvec{r}_j$
($j=1,\ldots,n*N$).
This ensemble is driven by an external beam with electric field
$\spvec{E}_{\rm in}(\spvec{r},t)$ and magnetic field $\spvec{B}_{\rm
  in}(\spvec{r},t)$ whose frequency components have wavelengths much
larger than the spatial extent of the meta-atoms.
We assume that meta-atom $j$ supports a single eigenmode of
current oscillation whose state can be described by a single dynamic
variable $Q_j(t)$ with units of charge and whose spatial profile is
described by time-independent functions $\spvec{p}_j(\spvec{r})$ and
$\spvec{w}_j(\spvec{r})$.
These mode functions are defined such that the polarization and
magnetization densities associated with atom $j$ are
$\spvec{P}_j(\rv,t) = Q_j(t) \spvec{p}_j(\spvec{r})$ and
$\spvec{M}_j(\rv,t) = I_j(t) \spvec{w}_j(\spvec{r})$ respectively,
where $I_j(t) = dQ_j/dt$ is the current.
We consider the Power-Zienau-Woolley Lagrangian in the Coulomb gauge
for the coupled system of the EM field and matter (described by an
arbitrary dynamic polarization and magnetization
densities)~\cite{CohenT}
\begin{equation}
  \label{eq:Lagrange2}
  \mathcal{L} = \mathcal{K} +\mathcal{L}_{\rm em}+ V_{\rm coul} +
  \sum_j \left[ Q_j(t) \mathcal{E}_j +   I_j(t) \Phi_j \right]\,,
\end{equation}
where $\mathcal{L}_{\rm em}$ is the Lagrangian for the free EM field,
$V_{\rm coul}$ is the instantaneous Coulomb interaction,
$\mathcal{E}_j(t) \equiv \int d^3r\, \spvec{E}(\spvec{r},t) \cdot
\spvec{p}_j(\spvec{r})$ is an electromotive force on the circuit $j$
and $\Phi_j(t) \equiv \int d^3r\, \spvec{B}(\spvec{r},t) \cdot
\spvec{w}_j(\spvec{r})$ is an effective field flux through circuit
$j$, $\mathcal{K} = \sum_j l I_j^2/2$, and $l$ is a phenomenological
constant representing an inertial inductance which is present in the
absence of any EM field interactions.
This inertial inductance may result, e.g., from the effective mass of
the charge carriers or surface plasmons as they  move through the
circuit.

We now derive a set of coupled equations governing the excitation of
the current oscillations driven by a near resonant field.
Following the procedure in Ref.~\cite{CohenT}, we find the system
Hamiltonian in terms of the dynamic variables $Q_j$ and their conjugate
momenta $\phi_j = lI_j + \Phi_j$ and arrive at the equations of motion $\dot{Q}_j(t) = (\phi_j(t)-\Phi_j(t))/l$ and $\dot{\phi}_j(t) = \mathcal{E}_j(t)$.
In order to integrate the scattered fields we express the positive
frequency components of the electric displacement and magnetic
induction in terms of the normal modes $a_{q}$, $\Dv^+ (\rv)=\sum_q
\pol_q \xi_q  a_q e^{i\qv\cdot\rv}$ and $\Bv^+
(\rv)=\sqrt{\mu_0/\epsilon_0} \sum_q \unitvec{k}\times\pol_q \xi_q a_q
e^{i\qv\cdot\rv}$, where $\Dv(\rv)=\Dv^+(\rv)+\Dv^-(\rv)$,
$(\Dv^+)^*=\Dv^-$, and $q$ refers to both wavevector $\qv$ and the
transverse polarization $\pol_q$.
We obtain the equations of motion for $a_q$ from \eq{eq:Lagrange2}.
The procedure resembles derivation for the expressions of the
scattered light in an atomic
vapor~\cite{RuostekoskiJavanainenPRA1997L} and full details will be
given elsewhere~\cite{JenkinsLongPRB}.
In the limit that the size of the circuit elements is much smaller
than the wavelength of light, we treat meta-atoms as radiating dipoles
and ignore the higher-order multipole-field interactions.
The electric and magnetic dipole moment for the $j^\textrm{th}$
meta-atom are $\spvec{d}_j = Q_j h_j \unitvec{d}_j$ and $\spvec{m}_j = I_j
A_j\unitvec{m}_j$ respectively, where $h_j$ and $A_j$ have units of
length and area, defined such that $h_j \unitvec{d}_j = \int
d^3r \spvec{p}_j(\spvec{r})$ and $A_j\unitvec{m}_j = \int d^3r
\spvec{w}_j(\spvec{r})$.
The total electric and magnetic fields are the sum of the incident
field and  the scattered field radiated by the meta-atoms
$\Ev(\rv,t)=\Ev_{\rm in}(\rv,t)+\Ev_{\rm S}(\rv,t)$ and
$\Hv(\rv,t)=\Hv_{\rm in}(\rv,t)+\Hv_{\rm S}(\rv,t)$, where $\Ev_{\rm
  S}(\rv,t)=\sum_j \Ev_{{\rm S},j}(\rv,t)$ and $\Ev_{{\rm
    S},j}(\rv,t)$ denotes the field emitted by the meta-atom $j$.
The Fourier components of $\Ev^+_{{\rm S}}(\rv,t)$ and $\Hv^+_{{\rm
    S}}(\rv,t)$ for the frequency $\Omega$ read ($k\equiv \Omega/ c$)
\begin{align}
  \spvec{E}^+_{{\rm S}}(\spvec{r},\Omega) &= \frac{k^3}{4\pi\epsilon_0} \int d^3 r' \,
  \Big[\sptensor{G}(\spvec{r} - \spvec{r}',k)
  \spvec{P}^+(\spvec{r}',\Omega)  \nonumber\\
  &  \qquad + \frac{1}{c} \sptensor{G}_\times (\spvec{r}-\spvec{r}',k)
  \spvec{M}^+(\spvec{r}',\Omega) \Big] , \\
  \spvec{H}^+_{{\rm S}}(\spvec{r},\Omega) &=
  \frac{k^3}{4\pi} \int d^3 r' \,
  \Big[\sptensor{G}(\spvec{r} - \spvec{r}',k)
  \spvec{M}(\spvec{r}',\Omega) \nonumber\\
  &  \qquad - c \sptensor{G}_\times
  (\spvec{r}-\spvec{r}',k) \spvec{P}^+(\spvec{r}',\Omega) \Big]\,,\\
  \label{eq:Green'sfunc}
  \sptensor{G}(\spvec{r},k) &= i\left[\frac{2}{3} \sptensor{1}
    h_0^{(1)}(kr) + \left(\frac{\spvec{r}\spvec{r}}{r^2} - \frac{1}{3}
      \sptensor{1} \right) h_2^{(1)}(kr) \right] \nonumber\\
  & - \frac{4\pi}{3}
  \sptensor{1} \delta(k\spvec{r})\,.
\end{align}
Here $\Pv^+(\rv)=\sum_j \Pv^+_j(\rv)$, where in the dipole
approximation $\Pv^+_j(\rv) \approx \spvec{d}^+_j\delta(\rv-\rv_j)$,
and $\sptensor{G}$ denotes the radiation kernel
representing the electric (magnetic) field emitted from an electric
(magnetic) dipole \cite{Jackson}, $h_n^{(1)}$ is the outward
propagating spherical Hankel function of order $n$, and
$\sptensor{G}_{\times}(\spvec{r},k)$ is the radiation kernel for the
magnetic (electric) field of an electric (magnetic) dipole source.
Specifically, $\sptensor{G}_{\times}$ acting on a vector $\spvec{v}$
yields ($\hat\rv\equiv \rv/r$)
\begin{equation}
  \label{eq:CrossGreen}
  \cdot \sptensor{G}_{\times} (\spvec{r},k)\cdot \spvec{v} =
  {e^{ikr}\over r}\big(1-{1\over ikr}\big) \,\hat\rv
  \times \spvec{v}\,\textrm{.}
\end{equation}

In our model the different circuit elements interact via dipole
radiation.
Each element $j$ also interacts with its own self-generated radiation,
resulting in an effective capacitance $C_j$
and self-inductance $L_j$ associated with the circuit current mode.
Calculation of $C_j$ and $L_j$ is analogous to the derivation of the
radiative linewidth of an atom and the values of $C_j$ and $L_j$
depend on the geometric structure of the mode functions $\spvec{p}_j$
and $\spvec{w}_j$~\cite{JenkinsLongPRB}.
Consequently, an isolated meta-atom exhibits a resonance frequency
$\omega_j \equiv 1/\sqrt{L_jC_j}$ and the charge oscillation results
in radiative damping rates for electric and magnetic dipoles,
$\Gamma_{E,j}(k) = h_j^2\omega_jC_j k^3/(6\pi\epsilon_0)$ and
$\Gamma_{M,j}(k) = \mu_0\omega_j A_j^2 k^3/(6\pi L_j)$, respectively.
We further simplify by neglecting the frequency dependence of
$\Gamma_{E/M,j}$ (the Markov approximation), the inertial inductance
(taking the limit $l/L_j \rightarrow 0$), and assume $\Gamma_{M,j} \ll
\omega_j$, so that a single meta-atom behaves as a simple LC circuit
with resonance frequency $\omega_j$ and a radiative dissipation rate
$\Gamma_{E,j} + \Gamma_{M,j}$.
In addition, meta-atoms exhibit nonradiative decay $\Gamma_O$, e.g.,
due to ohmic losses.

By including the inter-circuit interactions mediated by the field we
obtain the coupled equations of motion for the vector of dynamic
variables  $\colvec{Q}\equiv (Q_1,Q_2, \ldots, Q_{n*N})^T$ and
their conjugate momenta $\colvec{\phi} \equiv (\phi_1,\phi_2, \ldots,
\phi_{n*N})^T$
\begin{subequations}
  \label{eq:EqmApprox}
  \begin{align}
    -i\Omega\colvec{Q} =& \cmatrix{C}^{{1\over2}}
    \left[\left(\cmatrix{\omega}-i\cmatrix{\Gamma_M}\right)
      -3\cmatrix{\omega}^{{1\over2}}
      \cmatrix{\mathcal{G}_M}
      \cmatrix{\omega}^{-{1\over2}}/2\right]
    \cmatrix{L}^{-{1\over2}}\colvec{\phi}
    \nonumber \\
    & + 3\cmatrix{\mathcal{G}_\times}^T \colvec{Q}/2 -
    \indMat^{-1} \colvec{\Phi}_{\rm in}  \label{eq:QEqm}\\
    i\Omega\colvec{\phi} =& \cmatrix{L}^{{1\over2}}
    \left[\left(\cmatrix{\omega}-i\cmatrix{\Gamma_E}\right) -
      3\cmatrix{\omega}^{{1\over2}}
      \cmatrix{\mathcal{G}_E} \cmatrix{\omega}^{-{1\over2}}/2
    \right]
    \cmatrix{C}^{-{1\over2}}
    \colvec{Q} \nonumber\\
    &- 3
    \mathcal{G}_\times \colvec{\phi}/2 -
    \mathcal{E}_{\rm in}  \text{,} \label{eq:phiEqm}
  \end{align}
\end{subequations}
where we defined diagonal matrices $\cmatrix{C}_{j,j'} =
C_j\delta_{j,j'}$, $\cmatrix{L}_{j,j'} = L_j\delta_{j,j'}$,
$\cmatrix{\omega}_{j,j'} = \omega_j\delta_{j,j'}$,
$[\cmatrix{\Gamma_E]}_{j,j'} = \Gamma_{E,j}(k)\delta_{j,j'}$, and
$[\cmatrix{\Gamma_M}]_{j,j'} = \Gamma_{M,j} \delta_{j,j'}$.
The matrix $\indMat =
\cmatrix{L}^{1\over 2}[1+i\cmatrix{\Gamma}\cmatrix{\omega}^{-1} -(3/2)
\cmatrix{\omega}^{-{1\over 2}} \cmatrix{\mathcal{G}_M} \cmatrix{\omega}^{-{1\over
    2}} ]\cmatrix{L}^{1\over 2}$ appearing in Eq.~\eqref{eq:QEqm} is
the inductance matrix.
The driving terms are given in terms of the incident fields as
$\mathcal{E}_{{\rm in},j} = h_j \spvec{E}_{\rm in}(\spvec{r}_j,\Omega)
\cdot \unitvec{d}_j$ and $\Phi_{{\rm in},j} = A_j \spvec{B}_{\rm
  in}(\spvec{r}_j,\Omega) \cdot \unitvec{m}_j$.
The radiative interactions between the scatterers are described by
$\cmatrix{\mathcal{G}_{E,M,\times}}$.
These have vanishing diagonal elements and off-diagonal elements
($j\neq j'$) given by
\begin{align}
  \label{eq:GreensElecMat}
  \left[\cmatrix{\mathcal{G}_E}\right]_{j,j'} &= \unitvec{d}_j \cdot
  \sqrt{\cmatrix{\Gamma_E}} \sptensor{G}(\spvec{r}_j
  - \spvec{r}_{j'},k ) \sqrt{\cmatrix{\Gamma_E}}\cdot
  \unitvec{d}_{j'},\\
 \label{eq:GreenMagMat}
 \left[\cmatrix{\mathcal{G}_M}\right]_{j,j'} &= \unitvec{m}_j \cdot
 \sqrt{\cmatrix{\Gamma_M}}\sptensor{G}(\spvec{r}_j
 - \spvec{r}_{j'},k ) \sqrt{\cmatrix{\Gamma_M}}\cdot
 \unitvec{m}_{j'},\\
 \label{eq:emCrossMat}
  \left[\cmatrix{\mathcal{G}_\times} \right]_{j,j'} &= \unitvec{d}_j\cdot
  \sqrt{\cmatrix{\Gamma_E}}
  \sptensor{G}_\times(\spvec{r}_j -
  \spvec{r}_{j'},k)\sqrt{\cmatrix{\Gamma_M}} \cdot
  \unitvec{m}_{j'}\,.
\end{align}
The first lines of Eqs.~(\ref{eq:EqmApprox}a,b) are analogous to
frequency dependent inductance and capacitance matrices, respectively,
but due to the long-range interactions mediated by the radiation
field, these can substantially differ from the quasi-static
expressions for which $\cmatrix{\mathcal{G}_\times}$ is also absent.

We consider an incident EM field with a dominant frequency $\Omega$
and bandwidth $\delta\Omega$, and introduce the normal variables for
the meta-atoms  whose resonance frequencies are centered around
$\omega_0$
\begin{equation}
  \label{eq:normVar}
  b_j(t) = \frac{e^{i\omega_0 t}}{\sqrt{2\omega_j}} \left(\frac{Q_j}{\sqrt{C_j}}
  +i\frac{\phi_j}{\sqrt{L_j}}\right).
\end{equation}
When the EM field and the meta-atoms themselves have frequency components in a narrow
bandwidth about $\omega_0$, such that $\delta\Omega,
\Gamma_{E/M,j}, |\omega_j-\omega_0| \ll \omega_0$, we may
further simplify by substituting $\omega_0/c$ for $k$ in the
expressions for $\cmatrix{\Gamma_{E/M}}$ and
$\cmatrix{\mathcal{G}_{E/M/\times}}$.
Then, from Eqs.~\eqref{eq:EqmApprox} and \eqref{eq:normVar}, we arrive
at coupled equations for $b_j$'s.
We make the rotating wave approximation, in which we neglect terms
oscillating at high frequencies $\sim 2\omega_0$, and obtain the
following closed system for the current oscillation dynamics
\begin{equation}
  \label{eq:normVarEqmotion}
  \dot{\colvec{b}}(t) = \mathcal{C} \colvec{b}(t) + \colvec{F}(t) \text{,}
\end{equation}
where the coupling matrix and the driving are given by
\begin{align}
  \label{eq:CopulingMatC}
  \mathcal{C} =& -i\delta -\frac{\Gamma}{2}
  +{3\over4}\left(i\cmatrix{\mathcal{G}_E}  + i
    \cmatrix{\mathcal{G}_M}  + \cmatrix{\mathcal{G}_\times} +
    \cmatrix{\mathcal{G}_\times}^T\right),\\
  \label{eq:ForcingFunc}
  \colvec{F} =& i\frac{e^{i\omega_0 t}}{\sqrt{2\hbar\cmatrix{\omega}}}
  \left(\frac{1}{\sqrt{\cmatrix{L}}} \mathcal{E}_{\rm in}(t) +
    i\frac{1}{\sqrt{\cmatrix{C}}} \indMat^{-1}\colvec{\Phi}_{\rm in}(t)\right)\,.
\end{align}
Here the detuning $\delta\equiv \cmatrix{\omega} - \omega_0$ and
$\Gamma\equiv \cmatrix{\Gamma_E} + \cmatrix{\Gamma_M} +
\cmatrix{\Gamma_O}$, where we phenomenologically account for
ohmic losses through $\Gamma_O$.
In the limits we've considered, there exist as many distinct
collective eigenmodes of oscillation as there are meta-atoms in the
system.
Each collective mode corresponds to an eigenvector of the matrix
$\mathcal{C}$ and has a distinct resonance frequency and decay rate,
determined by the imaginary and real parts of the corresponding eigenvalue.
Eqs.~(\ref{eq:normVarEqmotion}-\ref{eq:ForcingFunc}) resemble the set
of equations describing a cooperative response of atomic gases to
resonant light in which case the coupling is due to electric dipole
radiation alone \cite{MoriceEtAlPRA1995,RuostekoskiJavanainenPRA1997L}.

The crucial component of the strong cooperative response of
closely-spaced scatterers are \emph{recurrent} scattering events
\cite{Lagendijk,RuostekoskiJavanainenPRA1997L} -- in which a wave is
scattered more than once by the same dipole.
Such processes cannot generally be modeled by the continuous medium
electrodynamics, necessitating the meta-atoms to be treated as
discrete scatterers.
An approximate calculation of local field corrections in a
magnetodielectric medium of discrete scatterers was performed in
Ref.~\cite{Kastel07} where the translational symmetry of an infinite
lattice simplifies the response.

We next apply the formalism to an array of ASRs.
A single ASR (metamolecule) consists of two separate coplanar circular
arcs (meta-atoms), labeled by $j\in\{l,r\}$ and separated by
$\spvec{a}\equiv \spvec{r}_r-\spvec{r}_l$.
The current oscillations in each meta-atom produce electric dipoles
with orientation $\unitvec{d}$ ($\unitvec{d}\perp\unitvec{a}$) and
magnetic dipoles with opposite orientations $\unitvec{m}_r =
-\unitvec{m}_l$ ($\unitvec{m}_r\perp\unitvec{a},\unitvec{d}$).
Each element of the ASR possesses decay rates $\Gamma_{E/M/O,j} =
\Gamma_{E/M/O}$.
An asymmetry between the rings, e.g., resulting from a difference in
arc length, manifests itself as a difference in resonance
frequencies with $\omega_r = \omega_0 + \delta\omega$ and $\omega_l = \omega_0 - \delta\omega$.
The interaction matrix $\mathcal{C}$ for a single ASR  is thus given by
\begin{equation}
  \label{eq:SplitRing}
  \left(
    \begin{array}{cc}
      -i\delta\omega -\Gamma/2 & 3 i
      \tilde\Gamma G/4 - 3\bar{\Gamma} S/4
      \\
      \frac{3}{4} i
      \tilde\Gamma G - 3\bar{\Gamma} S/4 &
      i\delta\omega -\Gamma/2
    \end{array}
  \right)
\end{equation}
where $\bar{\Gamma} \equiv \sqrt{\Gamma_E \Gamma_M}$, $\tilde\Gamma\equiv \Gamma_E - \Gamma_M$,
$G \equiv \unitvec{d}\cdot \sptensor{G}(\spvec{a},k) \cdot \unitvec{d}$, and $S
\equiv \unitvec{d} \cdot \sptensor{G}_\times (\spvec{a},k) \cdot
\unitvec{m}_r$ (which is here assumed to be real).
To analyze the collective modes of the ASR, we consider the dynamics
of symmetric $c_+$ and antisymmetric $c_-$ modes of oscillation
defined by $c_\pm \equiv (b_r \pm b_l)/\sqrt{2}$ that represent the eigenmodes of the ASR
in the absence of asymmetry $\delta\omega=0$,
From Eqs.~\eqref{eq:CopulingMatC} and \eqref{eq:SplitRing}, one finds
$\dot{c}_\pm =
(-\gamma_\pm /2 \mp i\Delta)c_\pm - i\delta\omega c_\mp + (F_r \pm
F_l)/\sqrt{2}$, where the decay rates  $\gamma_\pm = \Gamma_E [1 \pm 3
\mathrm{Im}(G)/2] + \Gamma_M[1 \mp 3\mathrm{Im}(G)/2] + \bar{\Gamma}
\mathrm{Re} (S) + \Gamma_O$, and the frequency shift $\Delta =
-3\mathrm{Re}(G) \tilde\Gamma /2 - 3\bar{\Gamma} \mathrm{Im} (S)/4$.
Excitations of these modes possess respective net electric and
magnetic dipoles and will thus be referred to electric and magnetic
dipole excitations. The split ring asymmetry $\delta\omega\ne0$ introduces an
EM coupling between these modes: the incident field with
$\spvec{E}_{\rm in} \parallel \unitvec{d}$ and $\spvec{B}_{\rm in} \perp
\unitvec{m}$ only excites the electric mode when
$\delta\omega =0$, but for $\delta\omega \ne 0$ it
can resonantly pump the anti-symmetric magnetic mode.

As an example of a collective response of a metamaterial array,
we study an ensemble of identical ASR metamolecules (with
$\spvec{a}=a \unitvec{e}_x$ and $\spvec{d}=d \unitvec{e}_y$) arranged
in a 2D square lattice within a circle of radius $r_c$ and with
lattice spacing $u$ and lattice vectors $\spvec{u}_1=u\unitvec{e}_x$
and $\spvec{u}_2 =u \unitvec{e}_y$.
The sample is illuminated by a cw plane wave
$\Ev_{\rm in}^+(\rv)= \half {\cal E} \unitvec{e}_y e^{i\kv \cdot \rv}$
with $\spvec{k} =k \unitvec{e}_z$, coupling to the electric dipole
moments of the ASRs.
A transmission resonance through such a sheet was
experimentally measured in Ref.~\cite{FedotovEtAlPRL2010} where
the number of active ASRs $N$ was controlled by decoupling
the ASRs with $r\agt r_c$ from the rest of the system with
approximately circular shaped metal masks with varying radii $r_c$.
The resonance quality factor increased with the total number of active ASRs,
saturating at about $N=700$.

\begin{figure}
  \centering
  \includegraphics{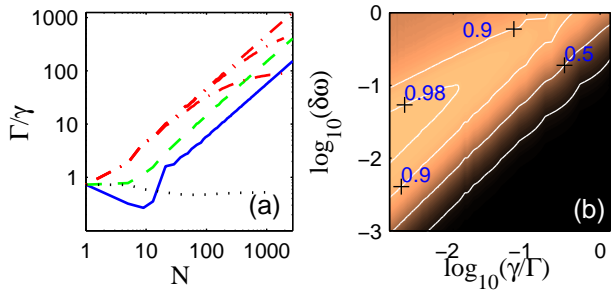}
  \caption{(Color online) The cooperative metamaterial response
    displaying collective resonance narrowing. (a) The resonance
    linewidth $\gamma$ of the collective magnetic mode $\colvec{v}_m$
    in the units of an isolated meta-atom linewidth
    $\Gamma$ (for $\Gamma_E = \Gamma_M$) as a function of the number
    of metamolecules $N$; the lattice  spacings $u=1/4\lambda$ (solid
    line), $3/8\lambda$  (dashed line),  $1/2\lambda$ (dot-dashed
    line), and $3/2\lambda$  (dotted line).
    The intermediate dot-dashed line corresponds to an asymmetry
    $\delta\omega  = 0.1\Gamma$, while $\delta\omega=0$ for all other
    curves.
    The  lower dot-dashed line incorporates ohmic loss (thermal
    absorption) $\Gamma_O=0.01\Gamma$ with all other curves assuming
    zero absorption.
    (b) Shows the overlap $O_m(\colvec{b}_F)$ of  $\colvec{v}_m$ with the
    state $\colvec{b}_f$ excited by an incident field resonant on the
    mode $\colvec{v}_m$ (for $u=1/2\lambda$, and $\Gamma_O=0$). For $\gamma
    \ll 1$ there is a range of asymmetries $\delta\omega$ for
    which the incident field almost exclusively excites the mode
    $\colvec{v}_m$.
  }
  \label{fig:GammaVNandOverlap}
\end{figure}

The incident field drives all metamolecules uniformly, and therefore
is phase-matched to collective modes in which the
electric and/or magnetic dipoles oscillate in phase.
In the absence of a split-ring asymmetry, only modes involving oscillating
electric dipoles can be driven. These modes strongly emit into the $\pm\unitvec{e}_z$ directions
enhancing incident wave reflection.
The magnetic dipoles, on the other hand, dominantly radiate into EM field modes
within the ASR plane. This radiation may become trapped through
recurrent scattering processes in the array, representing modes with
suppressed emission rates and reflectance, resulting in a transmission resonance.
Consequently, we study the radiation properties of the
magnetic eigenmode $\colvec{v}_m$ of $\mathcal{C}$ which maximizes
the overlap $O_m(\colvec{b}_A) \equiv | \colvec{v}_m^T \colvec{b_A} |^2/
\sum_i | \colvec{v}_i^T \colvec{b_A} |^2$ with the pure magnetic excitation
$\colvec{b}_A=(1,-1,\ldots,1,-1)^T/\sqrt{2N}$.
We then show that the introduction of an asymmetry allows the
excitation of $\colvec{v}_m$ by the incident field.
This mode closely resembles that responsible for the experimentally
observed transmission resonance
\cite{FedotovEtAlPRL2010,FedotovEtAlPRL2007}.

Figure \ref{fig:GammaVNandOverlap}a shows dependence of the resonance
linewidth $\gamma$ of $\colvec{v}_m$
on the number of metamolecules $N$ for different lattice spacings
$u$.
In the absence of ohmic losses and for sufficiently small $u$ and $\delta\omega$,
$\gamma\propto 1/N$ for large $N$.
The split ring asymmetry only weakly affects $\colvec{v}_m$.
For $\delta\omega=0.01 \Gamma$,  the curves representing
$\gamma$ are indistinguishable from those for $\delta\omega=0$.
For the relatively large $\delta\omega=0.1 \Gamma$, however, 
$\gamma$ is increased for $N\agt200$.
The cooperative response is very sensitive to $u$, resulting in
different linewidth narrowing.
For a larger $u=3/2\lambda$, however, $\gamma$ is
insensitive to $N$ indicating the limit of independent scattering of isolated metamolecules 
and a diminished role of cooperative effects.

As with an isolated ASR, an asymmetry $\delta\omega\neq0$
generates an effective two-photon coupling between
\emph{collective} electric and magnetic modes.
This is analogous to electromagnetically induced transparency:
one leg of the transition is provided by the coupling of the incident
field to collective electric dipole excitations, while the other is provided by
the effective coupling introduced by the asymmetry.
We illustrate this in Fig.~\ref{fig:GammaVNandOverlap}(b),
showing the relative population $O_m(\colvec{b}_f)$ of the collective
magnetic mode $\colvec{v}_m$ produced by a resonant field as a
function of $\gamma$ and 
$\delta\omega$, where $\colvec{b}_f$ is the state induced by the
uniform driving resonant on $\colvec{v}_m$.
One sees that for $\gamma\ll\Gamma$ and $\delta\omega \agt \gamma$,
one can excite a state  in which
more than $98\%$ of the energy is in the target mode $\colvec{v}_m$.
For $\delta\omega \ll \gamma$, any excitation that ends up in
$\colvec{v}_m$ is radiated away before it can accumulate; the array
behaves as a collection of symmetric metamolecules.
For larger $\delta\omega$, the population of $\colvec{v}_m$ decreases
since the increased strength of the effective two-photon transition
begins to excite other modes with nearby resonance frequencies.
Although the density of modes which may be excited increases linearly
with $N$,  the corresponding reduction of $\gamma$
means that a smaller $\delta\omega$ is needed to excite the target
mode, and there is a range of asymmetries for which $\colvec{v}_m$ is
excited.
The  narrowing in $\gamma$ combined with the near exclusive excitation
of this mode implies that for larger arrays the radiation
from the sheet is suppressed and hence the transmission
enhanced as in Ref.~\cite{FedotovEtAlPRL2010}.
The observed saturation \cite{FedotovEtAlPRL2010}, however, may
result from a combination of a fixed $\delta\omega$, which leads to
the population of other modes for larger arrays, and ohmic losses in
the resonators which put a floor on the collective decay rate
$\gamma$.
Fig.~\ref{fig:GammaVNandOverlap}a shows that an ohmic
loss rate of $\Gamma/100$ results in the expected saturation of
quality factor with $N$.

In conclusion, we developed a computationally efficient formalism
describing the collective interactions between discrete resonators.
In principle, one can calculate the EM response also by having knowledge of the material comprising the
circuit elements and numerically solving Maxwell's equations with a numerical mesh small enough
to resolve the features of each meta-atom. This, however, becomes computationally intractable when
the system contains more than a few resonators. In ASRs  we showed how an asymmetry leads to
excitation of collective magnetic modes by a field which does not couple directly to ASR magnetic moments
and results in cooperative response exhibiting a dramatic resonance linewidth narrowing, consistently with experimental findings~\cite{FedotovEtAlPRL2010}.

We acknowledge financial support from the EPSRC and discussions with N.\ Papasimakis, V.\ Fedotov, and N.\ Zheludev.


\begin{thebibliography}{30}


\bibitem{wiersma} D. S. Wiersma {\it et al.}, Nature {\bf 390}, 671 (1997).

\bibitem{negativeindex} D.R. Smith {\it et al.}, Science {\bf 305}, 788 (2004).

\bibitem{FedotovEtAlPRL2010} V. A. Fedotov \textit{et al.}, Phys. Rev. Lett. \textbf{104}, 223901 (2010).

\bibitem{SentenacPRL2008} A. Sentenac and P. C. Chaumet, Phys. Rev. Lett. \textbf{101}, 013901 (2008);
F. Lemoult \textit{et al.}, {\it ibid.} \textbf{104}, 203901 (2010); T. S. Kao {\it et al.}, arXiv:1010.1854 (2010).

\bibitem{papasimakis2009} N. Papasimakis \textit{et al.}, Phys. Rev. B \textbf{80}, 041102(R) (2009).


\bibitem{CohenT} C. Cohen-Tannoudji {\it et al.}, {\it
Photons and Atoms } (John Wiley \& Sons, New York, 1989).


\bibitem{RuostekoskiJavanainenPRA1997L} J. Ruostekoski and J. Javanainen, Phys. Rev. A {\bf 55},
513 (1997); {\bf 56}, 2056 (1997); J. Javanainen {\it et al.}, Phys. Rev. A {\bf 59}, 649 (1999).

\bibitem{JenkinsLongPRB} S. D. Jenkins and J. Ruostekoski, in preparation.

\bibitem{Jackson} J.D. Jackson, {\it Classical Electrodynamics}, (John Wiley \& Sons, New York, 1998).

\bibitem{MoriceEtAlPRA1995} O. Morice {\it et al.}, Phys. Rev. A
{\bf 51}, 3896 (1995).

\bibitem{Lagendijk} A. Lagendijk and B. A. van Tiggelen, Phys. Rep. \textbf{270}, 143 (1996).

\bibitem{Kastel07} J. K\"astel {\it et al.}, \pra {\bf 76}, 062509 (2007).

\bibitem{FedotovEtAlPRL2007} V. A. Fedotov {\it et al.} \prl \textbf{99}, 147401 (2007).

\end{thebibliography}

\end{document}